\documentclass[twocolumn,prd,groupedaddress,nofootinbib]{revtex4-1}

\usepackage{amsfonts,amsmath,amssymb,mathrsfs}
\usepackage{hyperref}
\usepackage{color}
\usepackage{graphicx}  
\usepackage{dcolumn}   
\usepackage{bm}        
\usepackage[english]{babel}

\def\a{\alpha}

\def\r{\rho}
\def\s{\sigma}
\def\t{\tau}
\def\m{\mu}
\def\n{\nu}
\def\k{\kappa}
\def\th{\theta}
\def\g{\gamma}\def\G{\Gamma}
\def\L{t}\def\l{V}
\def\D{\Delta}
\def\la{\langle}
\def\ra{\rangle}
\def\o{\omega}\def\O{\Omega}
\def\d{\delta}
\def\p{\partial}

\def\oxthree{{\cal O}(x^3) }

\def\half{\textstyle{\frac{1}{2}}}

\def\bdoc{\begin{document}}
\def\edoc{\end{document}}
\def\bea{\begin{equation}}
\def\eea{\end{equation}}

\def\beq{\begin{eqnarray}}
\def\eeq{\end{eqnarray}}
\def\ben{\begin{enumerate}}
\def\een{\end{enumerate}}
\def\la{\langle}
\def\ra{\rangle}
\def\a{\alpha}
\def\g{\gamma}\def\G{\Gamma}
\def\d{\delta}\def\D{\Delta}
\def\e{\epsilon}
\def\z{\zeta}

\def\th{\theta}
\def\k{\kappa}
\def\l{t}
\def\m{\mu}
\def\n{\nu}
\def\o{\omega}
\def\p{\pi}
\def\r{\rho}
\def\s{\sigma}
\def\t{\tau}
\def\L{{\cal L}}
\def\S{\Sigma }
\def\gsim{\; \raisebox{-.8ex}{$\stackrel{\textstyle >}{\sim}$}\;}
\def\lsim{\; \raisebox{-.8ex}{$\stackrel{\textstyle <}{\sim}$}\;}
\def\gtrsim{\gsim}
\def\lessim{\lsim}
\def\loc{{\rm local}}
\def\vm{v_{\rm max}}
\def\bh{\bar{h}}
\def\del{\partial}
\def\nab{\nabla}
\def\half{{\textstyle{\frac{1}{2}}}}
\def\fourth{{\textstyle{\frac{1}{4}}}}

\def\bD{{\bf D}}
\def\bE{{\bf E}}
\def\bF{{\bf F}}
\def\bB{{\bf B}}
\def\bP{{\bf P}}
\def\bV{{\bf v}}
\def\bv{{\bf v}}
\def\bx{{\bf x}}
\def\by{{\bf y}}
\def\bz{{\bf z}}
\def\ba{{\bf a}}
\def\bd{{\bf d}}
\def\bs{{\bf s}}
\def\bn{{\bf n}}
\def\bp{{\bf p}}

\def\O{\Omega}

\def\br{{\bf r}}
\def\bnab{{\bf \nab}}

\def\tE{\tilde{E}}
\def\tL{\tilde{L}}
\def\Horava{Ho\v{r}ava }

\def\oxtwo{\mathscr{O}\left(x^2\right)}
\def\oxthree{\mathscr{O}\left(x^3\right)}
\def\oxfour{\mathscr{O}\left(x^4\right)}
\def\oxfive{\mathscr{O}\left(x^5\right)}
\def\LL{Lanczos-Lovelock}

\def\ph{\phantom}

\begin{document}
\title{
The holographic entropy increases in quadratic curvature gravity
}
\author{Srijit Bhattacharjee}\email{srijitb@iitgn.ac.in}

\author{Sudipta Sarkar}\email{sudiptas@iitgn.ac.in}
\affiliation{Indian Institute of Technology, Gandhinagar, Gujarat 382424, India }
\author{Aron C. Wall}\email{aroncwall@gmail.com}
\affiliation{School of Natural Sciences, Institute for Advanced Study, Princeton, New Jersey 08540, USA}
\date{\today}
\begin{abstract}
Standard methods for calculating the black hole entropy beyond general relativity are ambiguous when the horizon is nonstationary.  We fix these ambiguities in all quadratic curvature gravity theories, by demanding that the entropy be increasing at every time, for linear perturbations to a stationary black hole.  Our result matches with the entropy formula found previously in holographic entanglement entropy calculations. We explicitly calculate the entropy increase for Vaidya-like solutions in Ricci-tensor gravity to show that (unlike the Wald entropy) the holographic entropy obeys a Second Law.
\end{abstract}
\maketitle

Black holes in general relativity (GR) obey laws of mechanics that are reminiscent of thermodynamical systems. One of the most profound is the Second Law of black hole mechanics, which %
classically 
states that the area $A$ of the event horizons of black holes increases irreversibly \cite{Hawking2nd, Bek, Bardeen:1973gs}\footnote{The proof requires an assumption of cosmic censorship in addition to the null energy condition}.
Another relation, the ``First Law'' (really the Clausius relation) takes the form $dE = T dS$ where $T$ is proportional to the surface gravity $\kappa$ which gives the rate at which light rays exponentially peel off the event horizon, entropy $S$ is proportional to the area, and $E$ is the Killing energy as measured in a coordinate system which is co-rotating with the black hole. Also, the Generalized Second Law (GSL)  states that the \emph{sum} of the entropies of the horizon and the matter outside $S_\mathrm{gen} = S_\mathrm{horizon} + S_\mathrm{out}$ is increasing, although entropy can be transferred from one term to the other \cite{Bekenstein:1974ax, Wall:2009wm, Wall:2010cj, Wall:2011hj}.

Hawking's discovery that black holes radiate thermally at a temperature $T = (\hbar \kappa/ 2\pi) $ \cite{Hawking:1974sw} strongly suggests that this is more than just an analogy; somehow these macroscopic laws derived from general relativity are describing the thermodynamics of some unknown microstates of the black hole.  

Additional evidence for a deep connection comes from gauge-gravity duality, where a stationary black hole is conjectured to be dual to a nonconfined thermal state of an ordinary gauge field theory \cite{Maldacena:1997re, Witten:1998qj, Aharony:1999ti}.  In this case the entropy of the black hole is equal to the ordinary thermodynamical entropy of the thermal fields at the boundary.  Inspired by this relationship, Ryu and Takayanagi \cite{Ryu:2006bv} proposed that the entropy in an arbitrary region $R$ of the field theory is given by a minimal area codimension $2$ surface in the gravitational theory, anchored to the boundary of the region $R$.  (In non-static situations, it is necessary to generalize this to the surface which extremizes the area \cite{Hubeny:2007xt}.)  The entropy of a stationary black hole is then a special case, found by choosing $R$ to be the entire boundary so that the minimal area surface is at the black hole horizon.  Although much evidence was found to support of this ``holographic 
entanglement entropy'' conjecture \cite{Ryu:2006ef, Nishioka:2009un}, it was only recently given a derivation by Lewkowycz and Maldacena \cite{Lewkowycz:2013nqa}, who showed how to calculate the entropy of the boundary theory using a $1$-parameter family of consistent solutions in the gravitational theory, and related this to the minimal area surface in the interior.

But in order for this story to be truly consistent, it is necessary to be able to take into account corrections to the action of general relativity.  After all, quantum theory has other effects besides enabling Hawking radiation.  Another effect, coming from loop corrections, is to renormalize the gravitational action.  On general Wilsonian grounds, one expects the effective field theory at an energy scale $\Lambda$ to have corrections depending on the Riemann tensor and various of its derivatives, suppressed by various powers (or logs) of $(\Lambda/ E_{\mathrm{planck}})$ as determined by dimensional analysis.  (In the nonexact Wilsonian renormalization group, we need only worry about those couplings which are divergent in the field theory as $\Lambda \to +\infty$; in four dimensions the cosmological constant and $1/G$ are power-law divergent, while the quadratic curvature corrections are log divergent.) \footnote{See \cite{Barrella:2013wja, Faulkner:2013ana, Engelhardt:2014gca} for additional 
discussion of how this affects the holographic entanglement entropy.}  

Do these higher curvature corrections spoil the laws of black hole thermodynamics which were the foundation of this entire line of research?  This is a deep question which has been studied from many angles.

The case of $f(R)$ gravity can be seen to obey a Second Law by virtue of it's equivalence to GR plus a scalar field by a field redefinition \cite{Jacobson:1993vj, Jacobson:1995uq, Ford:2000xg}, but other cases have proven more difficult.  Even Lovelock gravity, the most general covariant metric theory with just two time derivatives in the equation of motion, appears to violate the Second Law when two black holes merge \cite{Sarkar:2010xp}.  However, if we restrict attention to linearized metric perturbations to stationary black holes with a regular bifurcation surface, then the Second Law has been shown for all Lagrangians of the form $f(\mathrm{Lovelock})$ \cite{Chatterjee:2011wj, Kolekar:2012tq, Sarkar:2013swa}.  Our goal is to study whether the linearized Second Law continues to hold in the simplest theory not yet covered, namely Ricci-tensor squared theory.

A recent argument shows that one cannot have a consistent holographic theory of quantum gravity in which the Gauss-Bonnet term is finite in the classical limit $\hbar \to 0$, unless there is also an infinite tower of higher spin fields as in string theory \cite{Camanho:2014apa}.  (This is stricter than some previous bounds on the magnitude of Gauss-Bonnet (or Lovelock) terms \cite{Brigante:2008gz, Hofman:2008ar,Hofman:2009ug, deBoer:2009pn, deBoer:2009gx, Camanho:2009vw, Camanho:2009hu, Camanho:2010ru}).  However, it must still be possible to treat the higher curvature terms in a perturbative way, as part of a consistent quantum gravity or string theory expansion.

We hypothesize that in such a consistent perturbative truncation, it is important for the laws of thermodynamics to hold for higher curvature theories at linear order in the metric perturbation, but that they should not necessarily at nonlinear order (except of course for nonlinear pure GR effects). The argument for this comes from considering the case of an adiabatic (reversible) change in a quantum black hole.  In a reversible process (e.g. feeding the black hole with radiation whose temperature is very close to the Hawking temperature, but not exactly the same) the change in the black hole entropy can be made arbitrarily small: $dS_\mathrm{gen}/dt \approx 0$.  In the semiclassical regime such adiabatic processes involve only very small corrections to the black hole metric, which may however be of any sign.  But some higher curvature effects will still matter even at linear order in the metric perturbation (as will be seen explicitly for Ricci-tensor gravity below).  No matter how small the effects due to 
higher curvature corrections, for sufficiently adiabatic processes it will be necessary to take them into account in order to keep $dS_\mathrm{gen}/dt \ge 0$.  But if the GSL holds for linear perturbations to the metric, \emph{a fortiori} a classical Second Law should hold at linear order as well.

In order to discuss the validity of the First or Second Law, the first thing we need to know is what entropy formula to use.  The Noether charge method \cite{Wald:1993nt, Iyer:1994ys} says that the entropy of a stationary black hole is given by the generator of the boost symmetry at the horizon.  This is given by the Wald entropy formula obtained by differentiating the Lagrangian ${\cal L}$ with respect to the Riemann tensor \cite{Wald:1993nt}:
\bea
S_W =  - 2 \pi \int \frac{ \partial {\cal L}}{\partial R_{abcd} }\epsilon_{ab} \epsilon_{cd} \sqrt{\gamma}\,d^{D-2} x,\label{Wald_Exp}
\eea
where $\epsilon_{ab}$ is the binormal on the horizon slice and $\sqrt{\gamma}$ the area element.

Unfortunately, as noted by Jacobson, Kang, and Myers (JKM), the Noether charge method is ambiguous when applied to a slice of a nonstationary black hole horizon \cite{Jacobson:1995uq,Iyer:1994ys}.  As such the Wald entropy formula is just one of several possible candidates for the entropy.  JKM identified three different types of ambiguities, but the only one which is relevant in this context gives us the freedom to add to the Wald entropy any term of the form $ X \cdot Y$ where $X$ and $Y$ are objects which transform nontrivially under a boost of the two normal directions, but which are boost-invariant in combination \cite{Sarkar:2013swa}.  For example, the product of two extrinsic curvature tensors $K_{ab(i)} K^{ab(i)}$ or $K_{(i)} K^{(i)}$ are of this form (and are also have the same scaling dimension as the quadratic curvature entropy).  It is therefore unclear from the Noether charge method what the coefficients of these terms are.  Any entropy in this class is known to obey the First Law of black hole 
mechanics when comparing nearby stationary solutions \cite{Iyer:1994ys, Jacobson:1995uq}.  It will also obey the ``physical process version'' of the First Law, 
\cite{Jacobson:1995uq, Gao:2001ut, Jacobson:2003wv} in which a black hole begins and ends in a stationary state, but in the middle one perturbs it with a stress-energy tensor $T_{ab}$ and considers the linearized response:
\begin{equation}
\Delta S = \int_{\lambda}^{+\infty} T_{\bar{k}\bar{k}}\, (\lambda' - \lambda) \sqrt{\gamma}\,d\lambda' d^{D-2} x
\end{equation}
where $\lambda$ is an affine parameter along the null directions of the horizon, and $\bar{k}^a$ is the associated null vector.  Our notation is $A_{ab} k^a k^b = A_{kk}$. However, \emph{not} all entropies in this class have the entropy increasing instantaneously so that $dS/d\lambda \ge 0$.  Some JKM ambiguities (such as those quadratic in $K_{ab}^{(i)}$) matter even at linear order.  {\it In order to prove the Second Law, we must resolve these ambiguities.} 

Fortunately, the ambiguities have recently been resolved for the holographic entanglement entropy using the Lewkowycz-Maldacena method.  In the present case of quadratic curvature gravity the answer was calculated by Fursaev, Patrushev and Solodukhin \cite{Fursaev:2013fta, Solodukhin:2008dh} and by Camps \cite{Camps:2013zua}.  Dong \cite{Dong:2013qoa} did the more general case of $f(\mathrm{Riemann})$ gravity; some corrections and extensions to higher derivatives are in \cite{Miao:2014nxa, Miao:2015iba, Huang:2015zua}.  For Lovelock gravity see \cite{ Hung:2011xb, Bhattacharyya:2013gra, Bhattacharyya:2013jma, Bhattacharyya:2014yga}.


This raises the question of whether the holographic entropy also obeys the classical Second Law. In this work, by examining Vaidya-like solutions in Ricci-tensor gravity, we find that the holographic entropy obeys a local linearized Second Law but that the Wald entropy does not, just as in the case of $f(\mathrm{Lovelock})$ gravity \cite{Sarkar:2013swa}. So, not only the net change of the holographic entropy is positive, but it is increasing at every cross section of the horizon.

Our result provides strong support in favor of the holographic entanglement entropy conjecture from the physics of black holes, and highlights the importance of holographic duality for understanding the microscopic description of horizon entropy.  For the extension of our results to general linearized perturbations in arbitrary higher curvature gravity theories (using an entropy formula which matches Dong \cite{Dong:2013qoa} whenever both are valid) see Wall \cite{Wall4th}.

We start with the most general second order higher curvature theory of gravity described by the Lagrangian,
\beq
{\cal L} = (1 / 16 \pi)\left( R +\,\alpha\, R^2 +\,\beta \, R_{ab}R^{ab} +\, \gamma\,{\cal L_{GB}}  \right) \label{Flag}
\label{Lag1}\eeq
where ${\cal L_{GB}} = R^2\,-\,4 R_{ab}^2\,+\,R_{abcd}^2$ is the Gauss- Bonnet combination. As discussed before, the classical second law has already been proven for higher curvature corrections which are squares of the Ricci scalar \cite{Jacobson:1995uq} and proof of quasi stationary second law for Gauss-Bonnet combination in $D$-dimension is given in \cite{Sarkar:2013swa}  Therefore, if we establish an increase theorem of horizon entropy for the Lagrangian (\ref{Lag1}) with $\alpha,\,\gamma=0$, this will automatically provide a proof of the classical second law for a generic second order higher curvature theory of gravity in any dimension. So, we consider the relevant case when the Lagrangian is given by
\beq
{\cal L} = (1 / 16 \pi) \left( R + \,\beta \, R_{ab} R^{ab} + {\cal L_\mathrm{mat}}\right). \label{Lagrangian}
\eeq
Where we have also included the matter Lagrangian ${\cal L_\mathrm{mat}}$ which obeys the null energy condition.  (At first order in $\beta$, the gravitational part of this action could be transformed into the Einstein-Hilbert action by applying the field redefinition $\delta g_{ab} = \beta \delta (R_{ab} - \frac{1}{D-2} g_{ab} R)$.   Na\"{i}vely this suggests that the increasing entropy is $1 - \beta R_{ab} \epsilon^{ab}$, the Wald entropy; yet this is not so, because the transformation would also modify $L_\mathrm{mat}$ so that it no longer obeys the null energy condition.)

The equation of motion is: $G_{ab}\,+\,\beta H_{ab}=8\pi T_{ab}$ with,
\beq
 H_{ab} = &\Box R_{ab}\,- \,\nabla_a\nabla_b R\,+\,2R_{acbd}R^{cd}\, \nonumber \\ &+ \,{1\over 2}g_{ab}\left(\Box R\, - \,R_{cd}R^{cd}\right).
 \eeq
We parameterize the event horizon by a non-affine parameter `$t$'. We construct a basis with the vector fields $\{k^a, l^a, e^{a}_{A}\}$, where $l^a$ is a second null vector such that $l^a k_a = -1$. The induced metric on any $ t =$ constant slice of the horizon is $\gamma_{ab} = g_{ab} + 2 k_{(a} l_{b)}$. The horizon binormals are then given by $\epsilon_{ab} = \left( k_a l_b - k_b l_a\right)$.

In principle, 
a general theory of gravity might not possess any black hole solution.  But note that, the usual Schwarzschild black hole in general relativity is still a solution of Eq. (\ref{Lagrangian}). We also expect to have a spherically symmetric black hole solution with in-falling matter by perturbing this Schwarzschild solution, as a counterpart of the Vaidya solution in general relativity. Although, we do not know the solution explicitly, it must be of the form:
\bea
 dS^2=-f(r,v)dv^2\,+\,2dvdr\,+\,r^2d\Omega_{D-2}^2,\label{vaidya}
 \eea
where $f(r,v)$ is some arbitrary function. 

This solution has an event horizon whose location $ r = r(v)$ can be obtained by solving the equation $ d r(v) / dv = f(r, v) / 2$ with appropriate boundary condition. 
In the stationary limit, the metric function $f(r,v)$ vanishes on the event horizon. We now consider the null generators of the event horizon. In non affine parametrization, this is given by $k^a= \left(\partial_t\right)^a = \{2, f(r,v), 0, 0,...0\} $. The second null vector field is $ l_a=\{-1/2, 0, 0, 0,...0\}
$. These vector fields have the properties: $k_a k^a= l_a l^a=0\,\,;\,\, k^a l_a =-1$ and the null generator $k^a$ obeys the geodesic equation $ k^a\nabla_a k^b= \kappa \, k^b$. For our solution, $ \kappa = f'(r,v)$. Note that, in the stationary limit,  $\kappa/ 2 $ will coincide with the surface gravity of the background stationary solution. The expansions of the null congruences generated by $k^a$ and $l^a$ are $\theta_k$ and $\theta_l$ respectively and $\theta_{k}$ vanishes for stationary horizons. 


We will test whether the expression of holographic entanglement entropy for general quadratic curvature gravity obeys a linear second law in these Vaidya solutions.
In fact, the high symmetry of this non stationary solution makes the calculation tractable and explicit so that we can distinguish among the various possible expressions of horizon entropy beyond general relativity.  Since we are working in the linear approximation, we neglect all terms which are higher order in the perturbation such as $\theta_k^{2}$ etc. 

Now, we write the entropy of the black hole solution as,
\bea
 S=(1/4)\int \left(1 + \rho \right)\, \sqrt{\gamma}\,d^{D-2}x, \label{entropyG}
 \eea
where we have defined $\rho (t)$ as the entropy density contribution from the higher curvature terms and for general relativity, $\rho = 0$. 
Of course this must return the Wald entropy expression in the stationary limit.  Using this expression for the entropy, we calculate the entropy change and define the change of entropy per unit area as a generalized expansion given by,
\bea
 \Theta = \frac{d \rho}{dt} + \theta_{k}  \left(1 + \rho \right).
 \eea
 Using this form and neglecting some explicit higher order terms in Raychaudhuri equation, we obtain an evolution equation of $\Theta$:
\bea
\frac{d \Theta}{d t} - \, \kappa \Theta= - 8 \pi \,T_{kk}\, +   \nabla_k \nabla_k \rho - \rho R_{kk}  + \beta H_{kk} \label{evolution},
\eea
where we have used the field equation.  Note that, the first term in the $r. h. s.$ of Eq.(\ref{evolution}) is linear in the perturbation. If the rest of the terms collectively are of higher order, we can also ignore them and obtain,
\bea
\frac{d \Theta}{d t} - \kappa \Theta= - 8 \pi \,T_{kk}.
\eea
Further, if the matter obeys null energy condition, the above equation implies $d \Theta / d t - \kappa \Theta < 0$, on every slice of the horizon. In the asymptotic future, the horizon again settles down to a stationary state, so we must have $\Theta \to 0$ in the future. Now, if $\Theta$ is negative on any slice, since $\kappa > 0$, we would have $d \Theta / d t < 0 $  and as a result we would never have $\Theta$ to be zero in future. This means $\Theta$ must be positive on every slice prior to the future and this establishes that the entropy in Eq.(\ref{entropyG}) obeys a local increase law. So, if we want to prove an increase theorem for quasi stationary perturbation for any entropy candidate, what we need to show is $E_{kk} = \nabla_k \nabla_k \rho - \rho R_{kk}  + \beta H_{kk}  = {\cal O}(\epsilon^2)$ where $\epsilon$ is some parameter signifying the perturbation. 

Now, we start with the expression for Wald entropy in Eq. (\ref{Wald_Exp}). For the Lagrangian in Eq. (\ref{Lagrangian}), the Wald entropy density is $\rho_W = -2 \beta \, R_{ab} k^a l^b$. Using this Wald entropy density we find,
\bea
 E_{kk}  =\,\frac{2\, \beta (D-2)^2}{r^2}\frac{d^2f}{dv^2}\,+ {\cal O}(\epsilon^2). \label{waldevolution}
\eea
The leading term in the $r.h.s$ of the above equation is linear in the perturbation, and need not have any specific sign. So, it is evident even from this simple example that the Wald entropy density does not satisfy the requirement needed for a local increase law!  
This is not very surprising since the derivation of the Wald entropy from the stationary comparison version of the first law requires strict stationarity and existence of a regular bifurcation surface. Hence, the Wald entropy expression in Eq. (\ref{Lagrangian}) need not be valid in a non-stationary regime and could be corrected by ambiguity terms which vanish in the stationary limit.

Next we write down the entropy expression in \cite{Fursaev:2013fta, Solodukhin:2008dh, Camps:2013zua, Dong:2013qoa} motivated from the holographic entanglement entropy proposal. To translate the result of \cite{Dong:2013qoa} into our notation, we convert to Lorentzian signature and introduce two orthogonal vectors $n^{(i)}_a, \,i=1,2$ along the two dimensional space orthogonal to the horizon slice. They can be expressed in terms of the null vectors $k^a$ and $l^a$ as $n^{(1)a}=(1/ \sqrt{2})(k^a\,+\,l^a)\,;\,\,n^{(2)a}=(1/ \sqrt{2})(k^a\,-\,l^a)$.
The extrinsic curvatures of the horizon cross section along $n^{(i) a }$ are defined as $K^{(i)}_{ab}=(1/2){\cal L}_{n^{(i)}}\gamma_{ab}$ and it's trace is $K^{(i)}=K^{(i)}_{ab}\gamma^{ab}$.
 
We now calculate the entanglement entropy for our Lagrangian in Eq. (\ref{Lagrangian}). This is given by the following expression \cite{Dong:2013qoa},
 \bea
 S_D = \frac{1}{4}\int \left( 1 + \rho_W - \,{\beta \over 2}K_{(i)}K^{(i)}\right)\, \sqrt{\gamma}\, d^{D-2}x.
 \eea
 Where $\rho_W$ is the Wald entropy density. A simple computation gives us $K_{(i)}K^{(i)}=-2\theta_k\theta_l$ and we obtain the holographic entropy density: $\rho_D =\rho_W + \,\beta\, \theta_{k}\theta_{l}$. For the metric in Eq.(\ref{vaidya}), this is $ \rho_D = -2\beta\, R_{kl}\,- \beta (D-2)^2  f(r,v) / 2 r^2 $.
Note that the holographic entropy and the Wald entropy differ at the linear order!  So, although both of these expressions coincide for a stationary black hole, their evolutions are different once we perturb the horizon. To see that the holographic entropy obeys a local increase law, we first notice,
\beq
\nabla_k \nabla_k \rho_W = \nabla_k \nabla_k \rho_D +  \frac{2\, \beta \, (D-2)^2}{r^2}\frac{d^2f}{dv^2} + {\cal O}(\epsilon^2) \qquad 
\eeq
As a result, for the holographic entropy we have, $E_{kk} = {\cal O}(\epsilon^2)$. The offending term in the $r.h.s.$ of Eq. (\ref {waldevolution}) is canceled due to the presence of the term `$\beta \,\theta_{k}\theta_{l}$' in the holographic entropy, which implies a classical Second Law for Vaidya-type solutions (or a GSL, if we apply the methods of \cite{Wall:2011hj} to higher curvature gravity, as done in \cite{Sarkar:2013swa}).

Because of our restriction to spherical symmetry, our calculation does not yet indicate whether or not there should also be a $\sigma_{ab(k)}\sigma^{ab}\! {}_{(l)}$ shear-squared term in the entropy formula.  To see that there should not be such a term, consider a nonspherically symmetric vacuum solution to Einstein's equations with $R_{ab} = 0$, which is necessarily also a solution to any action composed purely of Ricci tensors.  This class of solutions includes stationary black holes perturbed by linearized gravity waves.  A linearized gravity wave allows one to set $\sigma_{ab(k)} \ne 0$, while $\sigma_{ab(l)} \ne 0$ can be arranged by choosing a 
wiggly slice of the horizon.  Integrating the Raychaudhuri equation, we find that $\theta_{k}$ vanishes at linear order, so the sole contribution to the entropy would come from the shear-squared term.  This allows us to easily violate the Second Law unless we choose the coefficient of the shear-squared term in the 
entropy to be $0$.  But $\sigma_{ab(k)}\sigma^{ab}\! {}_{(l)}$ and $\theta_{k}\theta_{l}$ were the only possible JKM ambiguities with the same dimension as $R_{kl}$.  It follows that holographic entropy is the \emph{only} entropy in the JKM class which can possibly obey a linearized increase theorem.

To summarize, we used a Vaidya-like non stationary black hole solution in a Ricci square theory of gravity and showed that the expression of holographic entanglement obtained in \cite{Fursaev:2013fta, Solodukhin:2008dh, Camps:2013zua, Dong:2013qoa} obeys a classical second law for linearized perturbations.  Our result combined with \cite{Jacobson:1995uq} and \cite{Sarkar:2013swa} suggests that the correct choice for black hole entropy density for a general quadratic curvature gravity described by the Lagrangian in Eq.(\ref{Flag}) is indeed
\beq
s = \frac{1}{4} +  \frac{1}{2} \left[\alpha R -  \beta\left( R_{ab} k^a l^b -  \frac{1}{2} \theta_{k}\theta_{l} \right) + \gamma\,r\,\right],\,\,\,\,
\eeq
where $r = R_{abcd} \gamma^{ac} \gamma^{bd} - K_{ab(i)}K^{ab(i)} + K_{(i)}K^{(i)}$ is the \emph{intrinsic} Riemann scalar on the horizon slice.  On a non-equilibrium slice of the horizon this entropy expression differs from the result obtained from the Wald's formula in Eq.(\ref{Wald_Exp}), but agrees with the holographic result in \cite{Fursaev:2013fta, Solodukhin:2008dh, Camps:2013zua, Dong:2013qoa}.  For general higher curvature theories see \cite{Wall4th}.

An important limitation of our result is that we have considered only the effects of matter falling into a black hole, not linearized gravitons.  In addition to not being spherically symmetric, infalling gravitons would require an analysis of second-order metric perturbations, and might require $\alpha, \beta, \gamma$ to satisfy certain constraints.  Also, in general gravitons travel on a different light cone than the matter fields \cite{Aragone:1987jm, ChoquetBruhat:1988dw, Brigante:2008gz}.

It seems that somehow the validity of black hole thermodynamics is already encoded in the holographic principle; the holographic entanglement entropy satisfies the linearized second law while the Wald entropy does not.  But the exact relationship with the holographic entanglement entropy is not completely clear.  Since slices of future horizons are not in general extremal surfaces, the entropy that increases cannot be fine-grained entropy of any boundary region.  By analogy to ordinary statistical mechanics, one expects that the increasing entropy must correspond to some \emph{coarse-grained} entropy.  Some recent proposals along these lines involve the causal holographic information \cite{Freivogel:2013zta, Kelly:2013aja, Kelly:2014owa} or the differential entropy \cite{Balasubramanian:2013lsa, Myers:2014jia, Czech:2014wka}.  It would be interesting to derive an entropy formula in one of these frameworks in a way that makes the holographic reason for the increase theorem explicit.\\

{\small 
The research of SS is partially supported by IIT Gandhinagar start up grant no. IP/IITGN/PHY/SS/201415-12.  AW is supported by the Institute for Advanced Study.
}

\end{document}